# High-quality electrical transport using scalable CVD graphene

Sergio Pezzini,[1, 2, 6, *] Vaidotas Mišeikis,[1, 2] Simona Pace,[1, 2] Francesco Rossella,[3] Kenji Watanabe,[4] Takashi Taniguchi,[5] and Camilla Coletti[1, 2, *]

[1]*Center for Nanotechnology Innovation @NEST, Istituto Italiano di Tecnologia, Piazza San Silvestro 12, 56127 Pisa, Italy*

[2]*Graphene Labs, Istituto Italiano di Tecnologia, Via Morego 30, 16163 Genova, Italy*

[3]*NEST, Scuola Normale Superiore and Istituto Nanoscienze-CNR, Piazza San Silvestro 12, 56127 Pisa, Italy*

[4] *Research Center for Functional Materials, National Institute for Materials Science, 1-1 Namiki, Tsukuba, 305-0044, Japan*

[5] *International Center for Materials Nanoarchitectonics, National Institute for Materials Science, 1-1 Namiki, Tsukuba, 305-0044, Japan*

[6] *Present address: NEST, Istituto Nanoscienze-CNR and Scuola Normale Superiore, Piazza San Silvestro 12, Pisa 56127, Italy*

E-mail: sergio.pezzini@nano.cnr.it; camilla.coletti@iit.it

**Abstract**

Producing and manipulating graphene on fab-compatible scale, while maintaining its remarkable carrier mobility, is key to finalize its technological application. We show that a large-scale approach (chemical vapor deposition on Cu followed by polymer-mediated semi-dry transfer) yields single-layer graphene crystals fully comparable, in terms of electronic transport, to micro-mechanically exfoliated flakes. Hexagonal boron nitride is used to encapsulate the graphene crystals – without taking part to their detachment from the growth catalyst – and study their intrinsic properties in field-effect devices. At room temperature, the electron-phonon coupling sets the mobility to ~$1.3 \times 10^5$ cm$^2$V$^{-1}$s$^{-1}$ at ~$10^{11}$ cm$^{-2}$ concentration. At $T$ = 4.2 K, the mobility (> $6 \times 10^5$ cm$^2$V$^{-1}$s$^{-1}$ at ~$10^{11}$ cm$^{-2}$) is limited by the devices' physical edges, and charge fluctuations < $7 \times 10^9$ cm$^{-2}$ are detected. Under perpendicular magnetic fields, we observe early onset of Landau quantization ($B$ ~ 50 mT) and signatures of electronic correlation, including the fractional quantum Hall effect.

**1. Introduction**

The high carrier mobility ($\mu$) of graphene is a central parameter in its investigation as material of technological relevance [1, 2]. As a result of the combination between light effective mass [3, 4] and low coupling between charge carriers and phonons [5, 6], graphene is expected to exhibit $\mu$ in excess of $10^5$ cm$^2$V$^{-1}$s$^{-1}$ at room temperature, promising bulk-semiconductors-beating performances [5]. However, due to its surface-only structure, graphene's properties are unavoidably altered by the surrounding environment, which strongly limits $\mu$ even in graphite-derived samples (~$10^4$ cm$^2$V$^{-1}$s$^{-1}$ if placed on SiO$_2$/Si) [7, 8]. Dean *et al.* [9] have been the first to show that hexagonal boron nitride (hBN) flakes, exfoliated from bulk single crystals [10], act as an ideal substrate for graphene, minimally affecting the charge carriers' motion. This concept was then developed into a dry pick-up technique by Wang *et al.* [11], who provided the community with a vital tool for fundamental studies [12], and demonstrated $\mu$ values only limited by intrinsic factors: electron-phonon scattering at room $T$, device-size at low $T$. Currently, state-of-the-art devices have all-van-der-Walls structures including graphite gates [13], which minimize charge disorder and increase the experimental sensitivity on delicate phases [14]. Nonetheless, the carrier mobility does not vary significantly with respect to Si-gated hBN-encapsulated samples [11, 14], which keep representing a solid benchmark for graphene electrical performances.

Despite these outstanding results, transferring high-mobility graphene from research laboratories to industrial-scale applications is proving to be extremely challenging [15]. Among the numerous scalable synthesis strategies developed over the years [16], chemical vapor deposition (CVD) of graphene single crystals on metal catalysts, such as Cu [17], combines the equally-required high material yield and quality. While almost every study on CVD-grown graphene (CVD-G) includes an evaluation of the electrical transport properties after transfer to $SiO_2$/Si substrate, only a handful of them reports on its performances when integrated in heterostructures with hBN, which is, at present, the only strategy to assess accurately its electronic quality. Petrone *et al.* [18] observed $\mu > 5 \times 10^4$ $cm^2V^{-1}s^{-1}$ at $T = 1.6$ K in CVD-G isolated by selective-area Cu etching and transferred on top of hBN, firstly showing flake-competitive mobility. Few years later, Banszerus *et al.* [19] developed a technique for direct pick-up from Cu using hBN, which, although ensuring $\mu > 10^5$ $cm^2V^{-1}s^{-1}$ at low $T$ in CVD-G-based devices, lacks in terms of scalability due to the small lateral size (typically up to ~100 μm) of the exfoliated hBN flakes. The same approach was used to demonstrate high carrier mobility in continuous graphene films at low $T$ [20]. Recently, De Fazio *et al.* [21] have applied hBN encapsulation to single-crystalline CVD-G after Cu etching and wet transfer to $SiO_2$/Si – that is, using a scalable growth and transfer technique – reporting intermediate performances between Ref.[18] and Ref.[19]. In this work, we use a scalable growth and transfer protocol that we developed for the integration of graphene in optoelectronics [22], and demonstrate, via subsequent hBN encapsulation, that the electrical transport properties – both at room and cryogenic temperatures – are comparable to those offered by exfoliated graphene flakes in equivalent devices. On this basis, we can conclude that, at present, the primary limitation to a graphene-based technology does not reside in CVD-G growth and transfer, but in the inadequacy of the large-scale dielectric substrates and encapsulation strategies at disposal.

## 2. Results and discussion

We begin our study with monolayer graphene single crystals grown on commercial Cu foils via low-pressure CVD (see Methods for details) [23], a technique that we optimized for the synthesis of large-scale arrays, intended to populate pre-patterned photonic circuits in a back-end-of-line approach [22]. We separate the crystals form the catalyst by electrochemical delamination in NaOH aqueous solution [24], while supporting them via a polymeric membrane, allowing easy handling and deterministic placing in dry conditions over arbitrary substrates [22, 25], $SiO_2$/Si in this case (see figure 1(a)). After cleaning in organic solvents, we obtain graphene crystals with a highly spatially-uniform Raman response, comparable to that of exfoliated flakes on the same substrate (negligible D peak, full width at half maximum of the 2D peak FWHM(2D) ~ 25-30 $cm^{-1}$, see black curve in figure 1(b)), indicating excellent material quality (see Supplementary information (SI), figure S1, for more details) [26, 27]. Our working hypothesis at this stage is the following: the electrical transport properties of devices based on these crystals are solely limited by the $SiO_2$/Si substrate, rather than being affected by growth and transfer. To verify this decisive point, we proceed with encapsulation in hBN flakes (figure 1(a)), using the pick-up-and-cleaning sequence described in Ref.[28]. To avoid the emergence of a moiré pattern and related effects on the electronic properties [12], straight edges of both hBN flakes are intentionally misaligned with respect to the hexagonally-shaped CVD-G crystals. Importantly, the pick-up is un-targeted, i.e., we do not make pre-characterization and selection of the CVD-G/hBN contact area, neither among, nor within the transferred crystals. An optical microscopy image of a typical hBN/CVD-G/hBN stack is shown in figure 1(c). As routinely observed in heterostructures of atomically thin crystals, a self-cleansing mechanism results in large blisters where the contaminants aggregate [29], separating flat areas where atomically sharp interfaces ensure the best electronic environment [30]. We conveniently individuate such areas by scanning Raman spectroscopy, obtaining spectra as the one shown in figure 1(b) (dark cyan curve) and false-color maps as the one in figure 1(d). The main parameter we monitor is FWHM(2D) (see SI, figure S2 for an analysis of other relevant Raman features) [27], which averages at 16-17 cm$^{-1}$ over the regions that we chose for fabrication of edge-contacted back-gated Hall bars (see Methods for details on the processing). Figure 1(e) shows the statistical distribution of FWHM(2D) measured over the active channels of three of such devices (D1-3, see SI figure S3 for optical microscopy and atomic force microscopy images), proving high spatial uniformity and no relevant differences in their response to Raman scattering.

In figure 2(a) we show the resistivity ($\rho$) of D1-3 as a function of the back-gate voltage ($V_{bg}$, applied to the underlying p-doped Si substrate) measured at room temperature and in vacuum (details on the measurement setup are given in Methods). The three devices show a narrow resistivity peak corresponding to charge neutrality, positioned at $V_{bg} \leq 0.5$ V, indicating minimal residual hole doping, with a maximum $\rho_0 = 1.10$-$1.25$ k$\Omega$. Away from the neutrality region, $\rho$ reduces to as low as 65 $\Omega$ (measured at $V_{bg} = -30$ V in D1). Although $V_{bg}$ can be applied to D3 only over a limited range (±2 V), due to an exponentially increasing leakage current, the narrow resistivity curve allows performing most of the relevant quantitative analysis also for this sample. In figure 2(b) we show a series of double-Log plots of the room $T$ conductivity $\sigma = 1/\rho$ as a function of the charge carrier density $n = C_{bg}(V_{bg} - V^0_{bg})/e$ (where $e$ is the electronic charge, $V^0_{bg}$ is the gate voltage at the charge neutrality point and the back-gate capacitance per unit area $C_{bg}$ is determined by low-field Hall effect measurements, see figure S4). We observe a linear Log($\sigma$) vs Log($n$) dependence, followed by a saturation when $n$ approaches ~$10^{10}$ cm$^{-2}$. By intersecting a fit to the linear part with a horizontal line set at the minimum conductivity $\sigma_0 = 1/\rho_0$, we estimate the charge carrier fluctuations $n^*$ for D1-3 to be within 3.2-4 × $10^{10}$ cm$^{-2}$. This range corresponds to the expected concentration of thermally excited carriers at room $T$ [31], implying that any disorder-induced inhomogeneity stands below this intrinsic broadening. For the sake of comparison, Ref.[21] reported $n^*$ in this order only at cryogenic temperatures. In figure 2(c) we show the mobility calculated according to the Drude model $\mu_D = \sigma/(ne)$ as a function of $n$ for D1-3 (the regions $|n| < n^*$ are excluded, since they correspond to a regime of coexisting electrons and holes [32]). As typically observed in high-quality hBN-encapsulated graphene [11], the curves show a plateau in the vicinity of $10^{11}$ cm$^{-2}$, where we observe a device-independent $\mu_D \sim$ 1.2-1.3 × $10^5$ cm$^2$V$^{-1}$s$^{-1}$; for comparison, note that Ref.[21] reported values lower by approximately a factor two at equal $n$ and $T$. With respect to CVD-G transferred on top of hBN [22], we observe a factor six increase in carrier mobility, highlighting the key role of dry hBN-encapsulation in probing the intrinsic electronic performance of the material. At higher carrier density, $\mu_D$ decreases due to electron-phonon scattering, as modelized by Hwang and Das Sarma [5], whose theoretical curve is plotted as a dashed line in figure 2(c) and represents a widely accepted upper bound for ideal environmentally-isolated single-layer graphene. In this sense, our data closely resemble the "textbook" ones reported by Wang *et al.* [11], which were obtained with exfoliated flakes. The recent findings on $\mu_D$ exceeding this limit in $WSe_2$-covered CVD-G [33] obviously cannot be compared to our results due to the different dielectric material employed; nevertheless, the reference hBN-encapsulated CVD-G devices reported there [33] show inferior $\mu_D$ with respect to D1-2 over the whole *n* range considered. Additionally, in SI (figure S5) we show that $\sigma(n)$ for D1-2 is well described by the relation $\sigma^{-1} = (ne\mu_L + \sigma_0)^{-1} + \rho_s$, where $\mu_L$ is a density-independent mobility (given by long-range scattering) and $\rho_s$ is a constant resistivity offset (due to short-range scattering) [7]. We obtain (independently on the device) $\mu_L = 1.5 \times 10^5$ cm$^2$V$^{-1}$s$^{-1}$ ($1.2 \times 10^5$ cm$^2$V$^{-1}$s$^{-1}$) for electrons (holes), roughly matching the $\mu_D$ plateau values, and $\rho_s = 46$ Ω (36 Ω) for electrons (holes), which corresponds to the expected magnitude of the resistivity due to electron-phonon coupling in ideal graphene [6].

Figure 2(d) shows $\rho(V_{bg})$ curves for D1-3 at $T = 4.2$ K. The resistivity peaks become extremely sharp and reach $\rho_0 = 3.8$-$4.2$ kΩ, while the smallest resistivity measured is 26 Ω (D1, $V_{bg} = 25.8$ V). The devices do not show diverging resistivity at the neutrality point, nor satellite peaks at large $V_{bg}$, thus confirming the rotational mismatch between CVD-G and the hBN flakes [12]. The cryogenic conditions suppress the thermally activated contributions and allow the observation of device-to-device variations in the width of the charge-neutrality peak, which reflect slight differences in the electrostatic disorder. To quantify this variability, we again employ double-Log $\sigma(n)$ plots (figure 2(e)) and estimate $n^* = 6.6 \times 10^9$ - $1.9 \times 10^{10}$ cm$^{-2}$ for the three devices. To the best of our knowledge, $n^*$ values in the $10^9$ cm$^{-2}$ range (obtained for D2-3), indicating extremely low potential fluctuations, have not been reported previously for CVD-G. Moreover, the device structure employed here is quite simple and does not include single-crystalline graphite gates that would further reduce $n^*$ by screening of remote disorder [13]. In figure 2(f) we plot $\mu_D$ for D1-3 as a function of $n$ (solid lines, excluding the regions $|n| < n^*$), together with $\mu_D = 4eW/(\pi h^2 n)^{1/2}$ (dashed lines, where $h$ is the Planck's constant), which is the expected carrier-dependent mobility for ballistic transport over distance $W$ [34], which we set equal to the devices' width (2.5 μm, 2 μm and 3 μm for D1, D2 and D3, respectively). This functional dependence captures the general behavior of the samples at large $n$, indicating that the devices' finite dimensions represent the primary limitation to the carriers' motion. Deviations and noisy features are ascribed to contacts-related fluctuations in the four-probe signals, due to finite width of the contact arms (see Figure S3) and high contact resistance (~kΩ). In the low-density range, we observe a slight electron-hole asymmetry, with the highest mobility reached at $|n| \sim 10^{11}$ cm$^{-2}$, where $\mu_D$ starts to approach the size-limited curves (see figure 2(f) inset). The peak values for D1-3 (averaged over a finite $n$ interval to account for fluctuations in the resistance signals) are in the range 4.1-6.6 × 10$^5$ cm$^2$V$^{-1}$s$^{-1}$ (2.1-3.6 × 10$^5$ cm$^2$V$^{-1}$s$^{-1}$) for electrons (holes), and identify our devices as the highest performing CVD-G-based to date.

In figure 3 we analyze our findings by studying the correlation between mobility and charge fluctuations. We do so by considering both $\mu_D(\sim 10^{11}$ cm$^{-2})$ (triangles) and the field effect mobility defined as $\mu_{FE} = (d\sigma/dn)/e$ (circles, where the slope $d\sigma/dn$ is obtained over the linear regions visible in figure 2(b), (e)), both for electrons and holes (filled and empty symbols). $\mu_D$ and $\mu_{FE}$ – values of mobility estimated via two different methods in a similar carrier density range – show a reasonable agreement over the whole plot, corroborating the discussion above, which is based on the Drude mobility ($\mu_D$). The data corresponding to measurements at 300 K collapse in a very narrow region, pointing at a *universal* behavior, i.e. determined solely by thermal broadening and insensitive to the sample details. When the devices are cooled to 4.2 K, the data show a more marked scattering, with D2-3 clearly positioning at higher $\mu$ and lower $n^*$ with respect to D1, reflecting the lower level of disorder. The overall behavior is well described by the relation $\mu^{-1} \propto n^*$ by Couto *et al.* [35] (dashed line in figure 3), as generally accepted for high-quality graphene on substrates.

To further support the observation of ultra-high carrier mobility in CVD-G, we measure the transport properties of D3 at $T = 0.3$ K, in the presence of a perpendicular magnetic field $B$. In figure 4(a) we show a false-color map of the longitudinal conductivity $\sigma_{xx} = \rho_{xx}/(\rho_{xx}^2 + \rho_{xy}^2)$ as a function of $V_{bg}$ and $B$ (up to 200 mT), where a typical fan of Landau levels (LL) can be appreciated. The condition to observe this phenomenology is governed by the competition between the cyclotron gap separating the LL $\Delta_c \sim 400 \sqrt{B(T)} \left( \sqrt{|N|} - \sqrt{|N-1|} \right) K$ (where $N$ is the LL index) and the disorder-induced level broadening $\Gamma = \hbar/2\tau_q$, where $\tau_q$ is the so-called quantum scattering time, which quantifies the carriers' scattering in presence of $B$. When a large enough field $B_{onset}$ is applied, $\Delta_c$ equals $\Gamma$ and the conductivity begins to oscillate, thus composing the fan-shaped diagram. In D3 we observe $B_{onset}$ as low as ~50 mT for the first oscillations at filling factor $\nu = nh/eB = \pm 2$, while it increases to ~100 mT for larger fillings (see light-red-to-white colored areas). In figure 4(b) we show $\tau_q$ as a function of $n$, extracted from the onset field of the oscillations (see SI, figure S6 for details on determination of $B_{onset}$). Close to charge neutrality, we observe $\tau_q < 0.1$ ps due to residual disorder in the $n^*$-region, with a marked growth to ~0.15 ps at higher density. To the best of our knowledge, the largest $\tau_q$ reported for graphene is 0.3 ps by Zeng *et al.* [36], who made use of exfoliated graphene, hBN-encapsulation and top and bottom graphite gates. The ×2 factor over our values (which, interestingly, corresponds to the difference in $n^*$ between D3 and their sample) can be mostly ascribed to the screening effect of the single-crystalline gates. Restricting our comparison to CVD-G only, Ref.[18] reported $B_{onset} = 400$ mT at $T = 1.6$ K, while Ref.[21] showed resolved LL at 1.8 T and 9

K. In figure 4(c) we show that the low-field oscillations lead to a fully developed quantum Hall effect already at $B$ = 200 mT, with zeroes in $\sigma_{xx}$ accompanied by quantized Hall conductivity $\sigma_{xy}$ = $\rho_{xy}/(\rho_{xx}^2 + \rho_{xy}^2)$ following the half-integer sequence of single-layer graphene [3, 4]. These observations prove ultra-low LL broadening and suggest the possibility of accessing correlation-driven phenomena making use of CVD-G. In figure 4(d) we show additional magnetotransport data on D3 (up to 12 T), where, starting from ~1 T, interaction-induced broken-symmetry states [37] are observed at $v$ = -3, -1 and 0. At charge neutrality ($v$ = 0) the sample becomes fully insulating for $B$ > 3 T (figure 4(e)), as expected for an interaction-induced spin-valley antiferromagnet [38]. Along filling factor $v$ = -1/3 (gray dashed line in figure 4(d)), we observe a zero-resistance region for $B \geq 8.5$ T, indicating a fractional quantum Hall (FQH) state [39, 40]. In addition to vanishing longitudinal resistance, quantum Hall states result in plateaus at $|v| \times e^2/h$ in the two-terminal conductance $G = I_{SD}/V_{SD}$ and $v \times e^2/h$ in $\sigma_{xy}$. Our measurements show plateau-like features, where $G \sim 1/3 \times e^2/h$ (figure 4(f)), while $\sigma_{xy}$ stands far from the expected value (figure 4(f), inset). In the SI we consider possible origins for this discrepancy and discuss how, in addition to higher magnetic field sources, more specialized device structures are likely needed for a thorough investigation of FQH in CVD-G. Nevertheless, FQH states require ultra-clean two-dimensional electronic systems [41], among which hBN-encapsulated CVD-G is to be included.

## 3. Conclusion

In conclusion, we presented unprecedented electrical transport performances for CVD-G. We synthesize and transfer single-layer graphene crystals using scalable approaches, and subsequently make use of hBN-encapsulation to investigate their intrinsic electronic response.

Our devices mimic the transport properties of equivalent samples (micrometer-sized and Si-gated) based on micro-mechanically exfoliated flakes, including room-$T$ mobility exceeding $10^5$ $cm^2V^{-1}s^{-1}$, onset of Landau quantization at ultra-low field, and signatures of FQH. While preparing this manuscript, we became aware that CVD-G detached from Cu via hBN-mediated dry pick-up, under high magnetic fields, shows similar evidence of FQH [42]. In addition to the high electronic quality of CVD-G, our work highlights a major weakness in the status of technological application of graphene, i.e. the lack of a large-scale analogue of hBN single-crystals. Although large-area few-layer hBN can be synthetized by CVD on metals, such material does not provide adequate environmental screening, resulting in poor graphene mobility if compared to devices employing exfoliated hBN flakes [43]. Regarding the use of CVD-G in fundamental research topics, this would require several improvements over the samples presented in this work. Apart from modifications in the device structure, increasing the size of the processable (bubble-free) regions within the hBN/CVD-G/hBN heterostructures is a clear priority. Engineering clean interfaces over large areas in CVD-G-based hetero-stacks is also of great technological relevance and might benefit from assembly in vacuum conditions [44], and post-assembly thermal and/or nano-mechanical treatment [31]. Moreover, large-area high-quality CVD-G carries considerable potential for the rising field of twisted bilayer graphene (TBG) [45], where the use of techniques such as micro-arpes and optical spectroscopies is currently limited due to the size of samples obtained via tear-and-stack of an individual exfoliated flake [46]. Since CVD-G single-crystals grown within the same mm-sized Cu grain are crystallographically aligned to each other [22], different crystals can be stacked with the required angular accuracy, facilitating large-scale studies of TBG. As an alternative, large-area TBG can be obtained via synthesis of CVD-bilayers with growth-controlled twist angle [25]. However, stabilizing small and accurate rotational shifts from the energetically favored Bernal stacking, as required for "magic-angle" TBG [45], poses significant challenges.

## 4. Methods

*Chemical vapor deposition:* We synthesize the graphene single crystals on electropolished Cu foils by CVD in a commercial reactor (Aixtron 4" BM-Pro), set at $p = 25$ mbar and $T = 1060$ °C. The Cu foil is annealed in Ar flow for 10 minutes at $T = 1060$ °C. The growth takes place for 15 minutes in 90% Ar, 10% $H_2$ and 0.1% $CH_4$. A quartz enclosure controls the gas flow on the sample [23], limiting the nucleation density.

*Raman spectroscopy:* We use scanning Raman spectroscopy to characterize the samples based on CVD-G. We employ a Renishaw InVia confocal spectrometer equipped with a 100× objective, with laser light at 532 nm wavelength, at ~ 1 mW laser power. The Si peak at 520 cm$^{-1}$ is used to calibrate the spectra.

*Device fabrication:* We process the hBN/CVD-G/hBN samples by e-beam lithography, reactive ion etching and thermal evaporation of metals. We first pattern the Hall bar mesa and etch the samples in $CF_4/O_2$. A second PMMA mask, followed by metal evaporation and liftoff, is used to define the electrical contacts (Cr/Au 5/70 nm), which connect to CVD-G via the exposed edges of the heterostructure [11]. The devices are glued on dual-in-line chip carriers using Ag conductive paste and wire-bonded with Al wires.

*Electrical transport measurements:* measurements at room $T$ are carried out in a vacuum chamber (base pressure $p \sim 10^{-5}$ mbar), with electrical connections to a dual-in-line holder. The low $T$ data are acquired in a $^4$He cryostat with superconducting coil for Hall effect measurements. In both cases, we use AC lock-in detection (13-17 Hz) in constant current configuration (10-100 nA). D3 is further tested in a $^3$He refrigerator providing additional cooling down to 0.3 K. The data in figure 4(d)-(f) are acquired in a constant voltage configuration ($V_{SD}$ = 100 µV), measuring, in addition to the longitudinal and Hall voltage drops, the source-drain current.

**Figures**

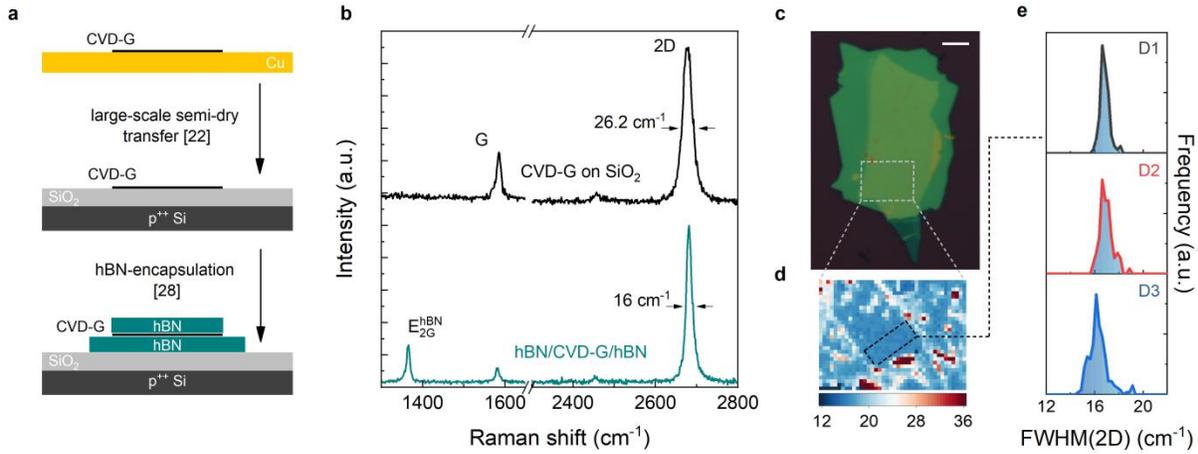

**Figure 1.** (a) Schematics of the samples' preparation: CVD-G single crystals are synthesized on Cu and transferred to SiO$_2$/Si; afterwards, they are "sandwiched" between hBN flakes by dry pick-up. (b) Representative Raman spectra of CVD-G on SiO$_2$/Si (black) and of hBN/CVD-G/hBN (dark cyan; the two spectra are normalized at the 2D peak maximum). FWHM(2D) is indicated for the two curves. (c) Optical microscopy image of a typical hBN/CVD-G/hBN sample (scale bar is 10 μm). The gray dashed rectangle marks the region over which we perform scanning Raman mapping. (d) False-color map of FWHM(2D) over the area indicated in (c). FWHM(2D) < 18 cm$^{-1}$ is measured over flat parts of the heterostructure (light blue), while interface bubbles give FWHM(2D) > 30 cm$^{-1}$ (dark red). The black dashed rectangle indicates the part of the sample employed for fabrication of device D1. (e) Statistical distribution of FWHM(2D) measured on the active channels of devices D1-3. The color scale is the same as in (d).

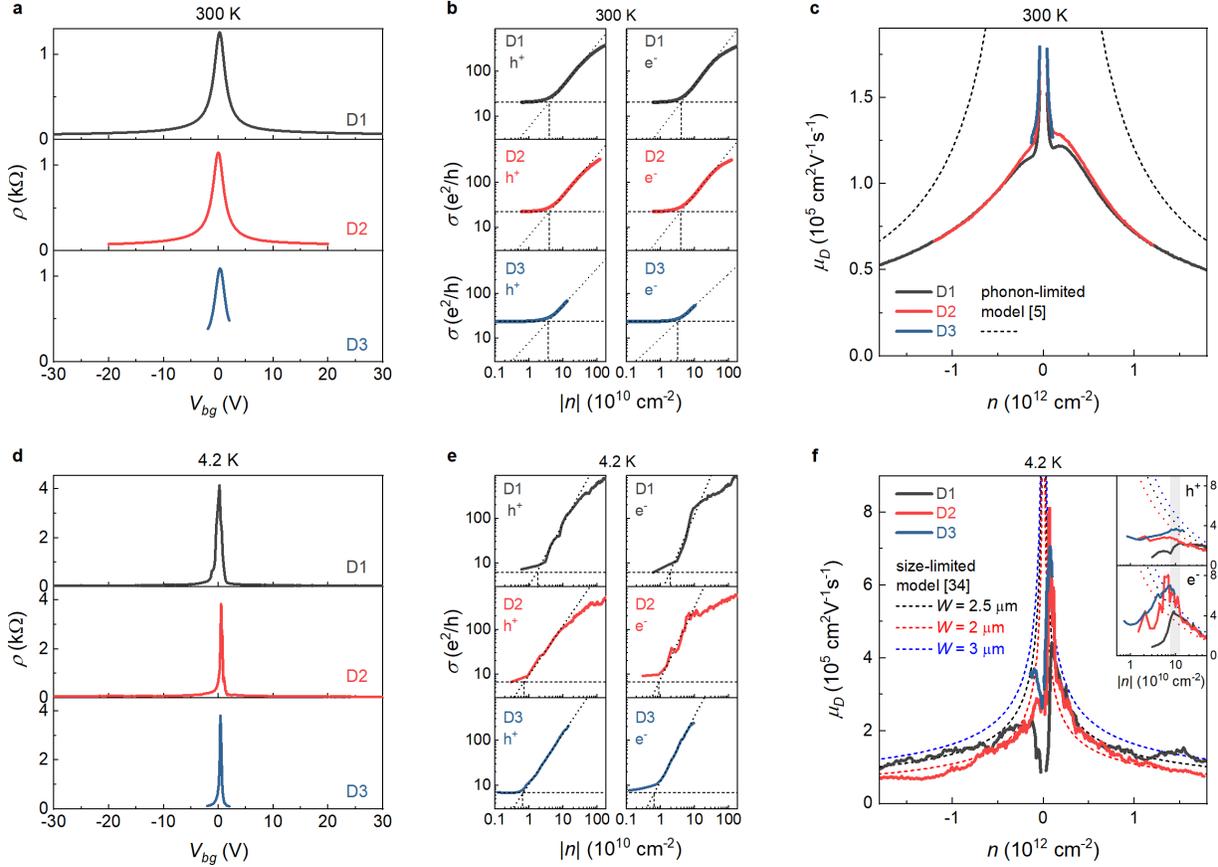

**Figure 2.** (a) $\rho$ as a function of $V_{bg}$ for devices D1-3, measured at room temperature ($T = 300$ K). (b) Double-Log $\sigma(n)$ plots obtained from the data in (a), both for hole (left panels) and electron doping (right panels). The horizontal dashed lines are set at the minimum conductivity for each device, the oblique ones are linear fits to Log($\sigma$)(Log($n$)) and the vertical ones connect the intersection of the first two to the x axis, determining the estimate of $n^*$. (c) $\mu_D$ as a function of $n$ for devices D1-3. The dashed lines are calculations from the model of Ref.[5], showing the theoretical concentration-dependent phonon-limited mobility of graphene at room temperature. (d) Same as (a) but measured at liquid helium temperature ($T = 4.2$ K). (e) Same as (b) but from the gate-dependent data at $T = 4.2$ K shown in (d). (f) Same as (c) but at $T = 4.2$ K. The dashed lines are calculations for density-dependent ballistic transport over distances up to the widths of the Hall bars [34]. The inset shows separate enlarged views of the main panel for hole and electron doping, with Log scale on the x axis, highlighting the peak region of $\mu_D$. The shaded rectangle indicates the averaging interval (0.75-1.25 × $10^{11}$ cm$^{-2}$) used to calculate the values of $\mu_D$ reported in the text and in figure 3.

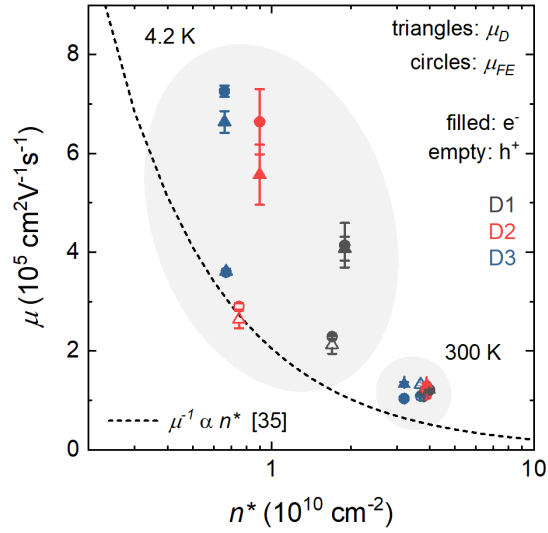

**Figure 3.** Mobility as a function of $n^*$, summarizing our results for D1-3, both at room and liquid helium temperature. The triangles are estimates from the Drude model at carrier concentration ~$10^{11}$ cm$^{-2}$ (the error bars are standard deviations over the averaging interval). The circles are field-effect mobility (the error bars are from the linear fits to $\sigma(n)$). The filled (empty) symbols are for negatively (positively) charged carriers. The dashed line is a model from Ref.[35], describing an inverse proportionality between $\mu$ and $n^*$.

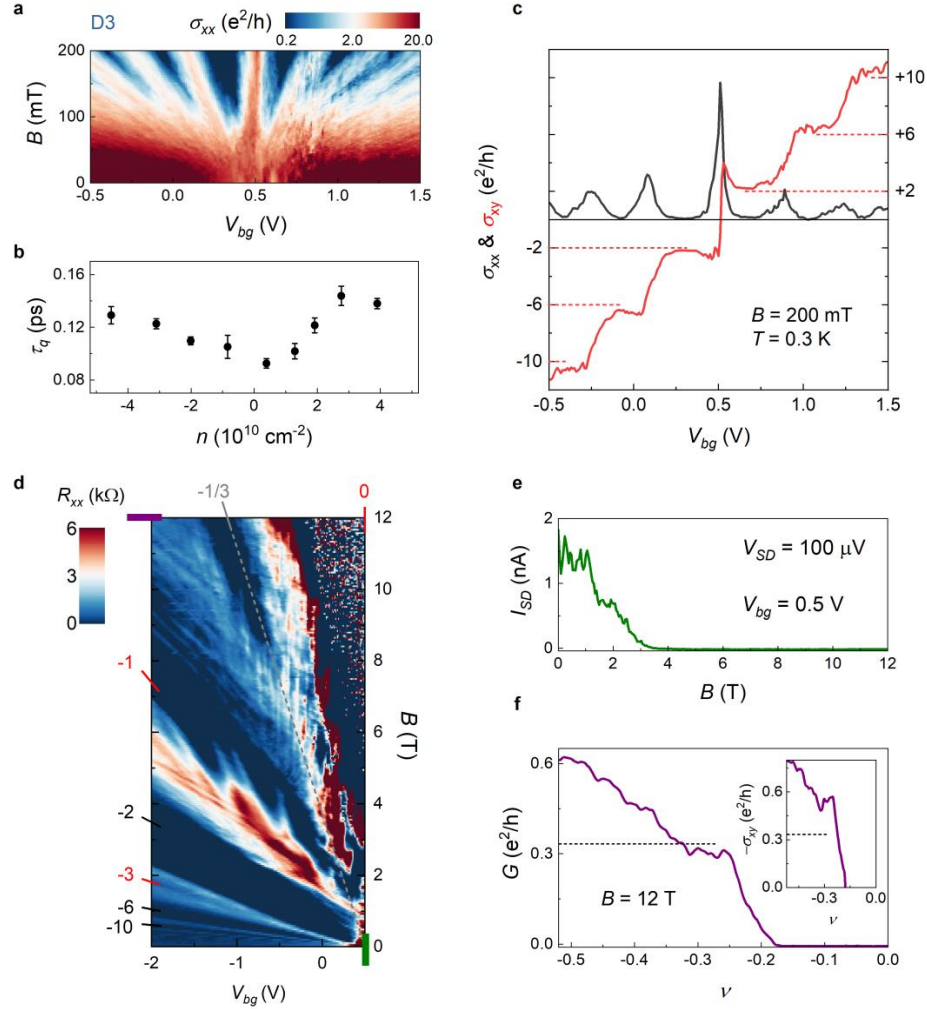

**Figure 4.** (a) False-color map of the longitudinal conductivity of D3 as a function of back-gate voltage and magnetic field. The logarithmic color scale is chosen to highlight the Landau level fan and the low-field onset of the oscillations. (b) Quantum scattering time as a function of the carrier density, obtained from the onset field as described in the text. (c) Longitudinal (black) and Hall conductivity (red) of D3 as a function of the back-gate voltage at fixed field $B = 200$ mT. The horizontal dashed red lines indicate the plateau values of the half-integer quantum Hall effect in graphene monolayer. (d) False-color map of the longitudinal resistance of device D3 as a function of back-gate voltage and magnetic field. The four-fold degenerate integer quantum Hall states (-2, -6, -10) are labelled in black, the broken-symmetry integer quantum Hall states

(0, -1, -3) in red, the -1/3 fractional state in gray. The gray dashed line indicates filling factor $v =$ -1/3. (e) Source-drain current as a function of the magnetic field, measured at charge neutrality (green mark in (d)). (f) Two-terminal conductance as a function of the filling factor, measured at the highest magnetic field at disposal, $B = 12$ T (purple mark in (d)). The curve is corrected by considering a contribution from contact resistance of 6.2 k$\Omega$, estimated from the offset in the conductance plateau value at $v = -2$, $B = 2$ T (increased contact resistance at 12 T likely accounts for the imperfect quantization). The inset shows the Hall conductivity under the same conditions. All the data are acquired at $T = 0.3$ K.

**Acknowledgements**

Growth of hexagonal boron nitride crystals was supported by the Elemental Strategy Initiative conducted by the MEXT, Japan, Grant Number JPMXP0112101001, JSPS KAKENHI Grant Numbers JP20H00354 and the CREST(JPMJCR15F3), JST. This project has received funding from the European Union's Horizon 2020 research and innovation programme Graphene Flagship under grant agreement No 785219 and No 881603.

# Supplementary information:

# High-quality electrical transport using scalable CVD graphene

Sergio Pezzini,[1,2,6,*] Vaidotas Mišeikis,[1,2] Simona Pace,[1,2] Francesco Rossella,[3] Kenji Watanabe,[4] Takashi Taniguchi,[5] and Camilla Coletti[1,2,*]

[1]*Center for Nanotechnology Innovation @NEST, Istituto Italiano di Tecnologia, Piazza San Silvestro 12, 56127 Pisa, Italy*

[2]*Graphene Labs, Istituto Italiano di Tecnologia, Via Morego 30, 16163 Genova, Italy*

[3]*NEST, Scuola Normale Superiore and Istituto Nanoscienze-CNR, Piazza San Silvestro 12, 56127 Pisa, Italy*

[4] *Research Center for Functional Materials, National Institute for Materials Science, 1-1 Namiki, Tsukuba, 305-0044, Japan*

[5] *International Center for Materials Nanoarchitectonics, National Institute for Materials Science, 1-1 Namiki, Tsukuba, 305-0044, Japan*

[6] *Present address: NEST, Istituto Nanoscienze-CNR and Scuola Normale Superiore, Piazza San Silvestro 12, Pisa 56127, Italy*

E-mail: sergio.pezzini@nano.cnr.it; camilla.coletti@iit.it

**Raman spectroscopy of CVD-G and hBN/CVD-G/hBN**

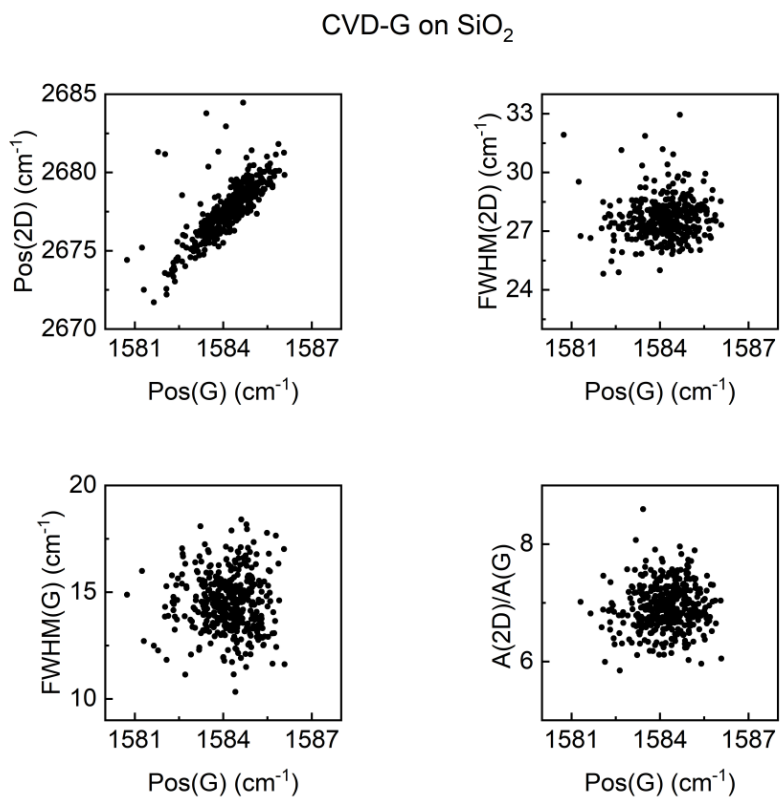

**Figure S1.** Raman correlation plots for CVD-G after transfer to $SiO_2$/Si.

In Figure S1 we show four plots correlating the main Raman parameters, measured on a CVD-G crystal from the same growth batch of the ones used for the devices discussed in the main text, after transfer to $SiO_2$/Si (200×200 µm$^2$ area). Pos(G) averages at 1584.2 cm$^{-1}$, indicating minimal uniaxial (biaxial) strain ~0.05% (0.02 %), calculated considering the influence of the finite doping level (see below) [S1]. The average FWHM(2D) is 27.7 cm$^{-1}$ and can be ascribed to strain fluctuations induced by the rough $SiO_2$ substrate within the size of the laser spot [25]. The combined average FWHM(G) = 14.5 cm$^{-1}$ and A(2D)/A(G) = 6.9 indicate hole doping ~100-150 meV [S2].

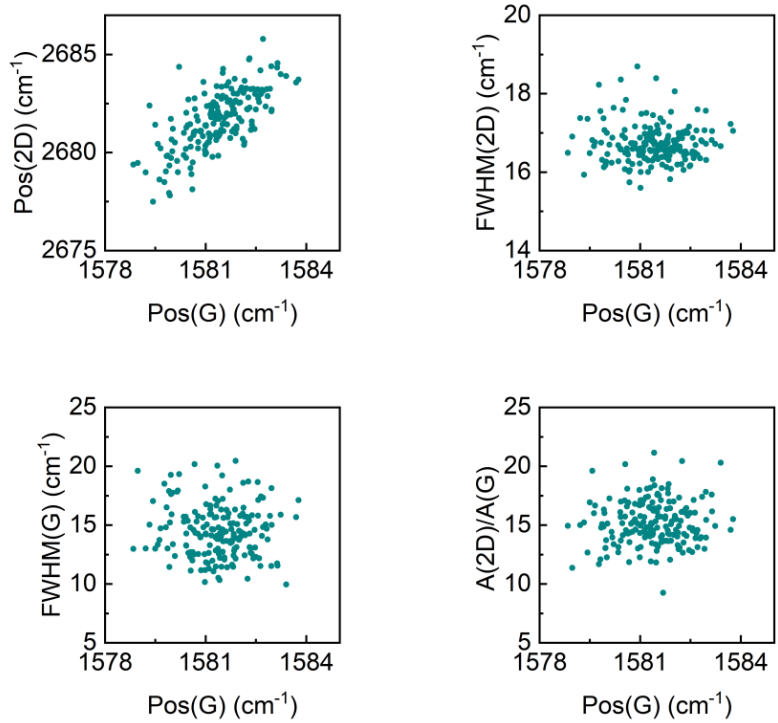

**Figure S2.** Raman correlation plots on a blister-free flat area of hBN/CVD-G/hBN (black dashed rectangle in main text Figure 1d). Device D1 is fabricated over this region.

The same kind of data for a bubble-free area of hBN/CVD-G/hBN (black dashed rectangle in main text Figure 1d) are shown in Figure S2. The average Pos(G) = 1581.4 cm$^{-1}$ is characteristic of pristine single-layer graphene [S3], while the slight blue-shift in Pos(2D) = 2681.7 cm$^{-1}$ can result from dielectric screening by hBN [S4]. The average FWHM(2D) = 16.7 cm$^{-1}$ indicates minimal strain fluctuations [25]. The average A(2D)/A(G) = 15.3 suggests a reduction of the carrier concentration to the undoped intrinsic limit [S2], which is corroborated by the field-effect electrical transport measurements (main text Figure 2).

**Atomic Force Microscopy of hBN/CVD-G/hBN devices**

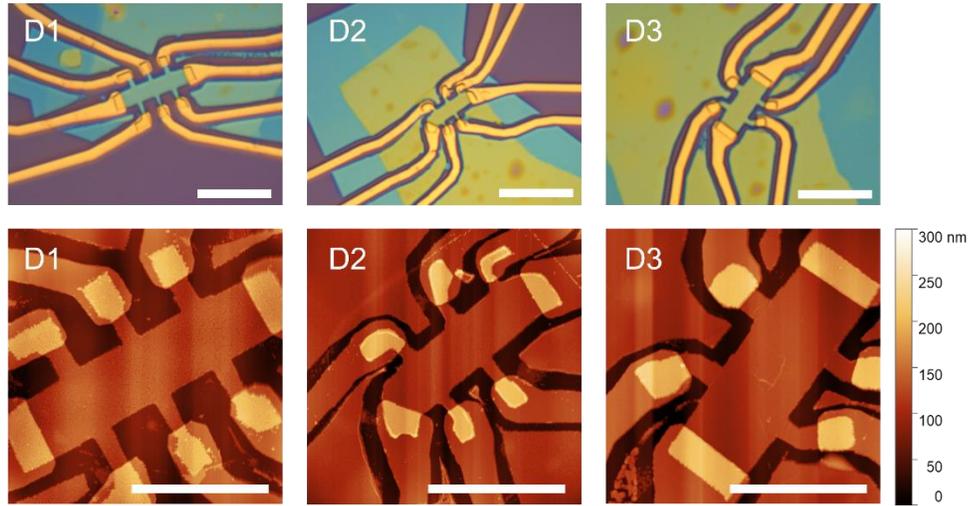

**Figure S3.** Upper panels: optical microscopy images of Hall bars D1-3 (scale bars are 10 μm). Lower panels: AFM images of the same devices (scale bars are 5 μm). The total thickness of the heterostructures is 70-80 nm (30-40 nm for each hBN flake). The geometrical factor for the contacts used in the resistivity measurements is $W/L = 1$ for the three samples. D1 shows larger surface roughness with respect to D2-3, due to polymer residuals after the etching step.

**Gate-dependent Hall effect and estimate of the gate capacitance**

To determine the back-gate capacitance per unit area ($C_{bg}$) of the devices, we perform Hall effect measurements as a function of the back-gate voltage in presence of $|B| = 0.5$ T, oriented perpendicular to the sample plane. We measure $R_{xy}(V_{bg})$ for positive and negative field orientation, and antisymmetrize the data according to $R_{xy} = (R_{xy}(+B) - R_{xy}(-B))/2$, to eliminate longitudinal components due to slight contact misalignment, obtaining curves as the one shown in Figure S4, left panel. Since the carrier concentration is given by $n = B/e \times 1/R_{xy} = C_{bg}(V_{bg} - 

$V^0_{bg}$)/$e$, $C_{bg}$ can be obtained from a linear fit to $1/R_{xy}$ as a function of ($V_{bg}$ - $V^0_{bg}$), as shown in Figure S4, right panel. For D2, we estimate $C_{bg}$ = 0.97 × 10$^{-8}$ F/cm$^2$, in agreement with the calculated capacitance per unit area of two in-series parallel plate capacitors made by the SiO$_2$ layer (thickness $d$ = 300 nm and dielectric constant $\varepsilon_r$ = 3.7) and the bottom hBN flake ($d$ = 30 nm and $\varepsilon_r$ = 3).

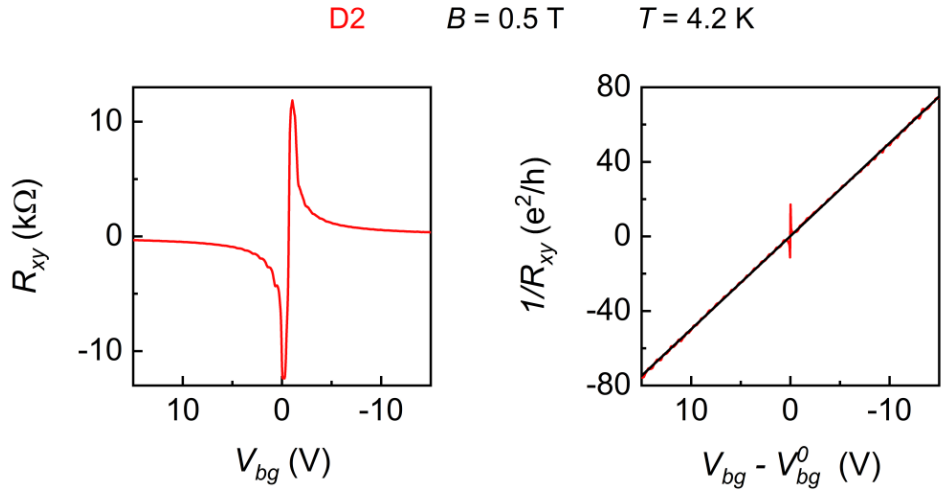

**Figure S4.** Left panel: field-symmetrized Hall resistance of device D2 as a function of the back-gate voltage, measured at |$B$| = 0.5 T and $T$ = 4.2 K. Right panel: inverse of the Hall resistance as a function of the back-gate voltage relative to the neutrality point (red curve). The black line is a linear fit.

**Room temperature conductivity for D1-2**

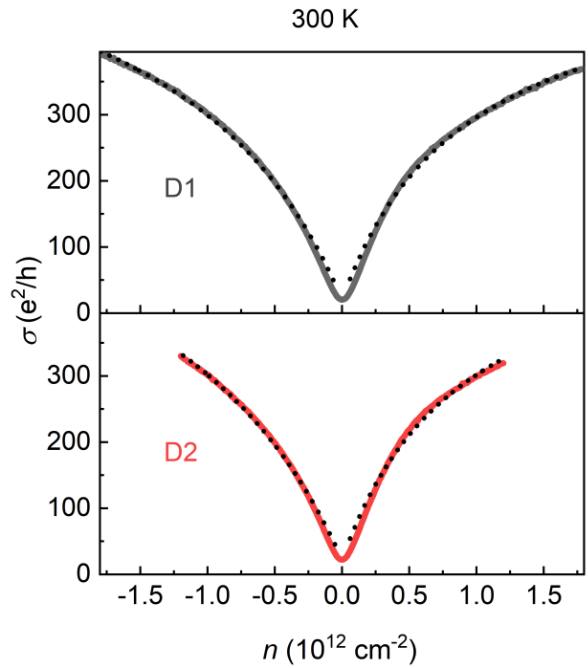

**Figure S5.** Room temperature conductivity as a function of carrier density for D1-2 (gray and red solid lines, respectively). The black dashed lines are fits to the data using the relation $\sigma^{-1} = (ne\mu_L + \sigma_0)^{-1} + \rho_s$ [7]. The fit is performed separately for holes and electrons and excludes the saturation region close to $n^*$. The fitting parameters $\mu_L$ and $\rho_s$ for the two devices are within 3% of the values given in the main text.

**Estimate of the Landau quantization onset**

In order to extract the quantum scattering time (data in Figure 4b), we quantify $B_{onset}$ for each oscillation of the low-field LL fan (Figure 4a). Here we use the $\nu = -6$ oscillation as an example.

We first inspect zoomed-in parts of the conductivity false-color map to estimate an onset region (see Figure S6, left). This provides us an interval of densities over which this specific oscillation starts to be observable, from -1.86 to -2.16 × 10$^{10}$ cm$^{-2}$ in this case. For each density measured in such an interval, we look at the longitudinal resistance as a function of the magnetic field (Figure S6, right, top panel) and identify the oscillation minimum corresponding to the filling factor of interest (-6 in this case). To quantitatively address the onset, we consider the first derivative of the resistance (bottom panel, calculated after smoothing the resistance signal) and identify the onset field as corresponding to the maximum negative point in d$R_{xx}$/d$B$ before the oscillation minimum. The onset field values are then averaged over the density interval to obtain $B_{onset}$, and $\tau_q$ is calculated.

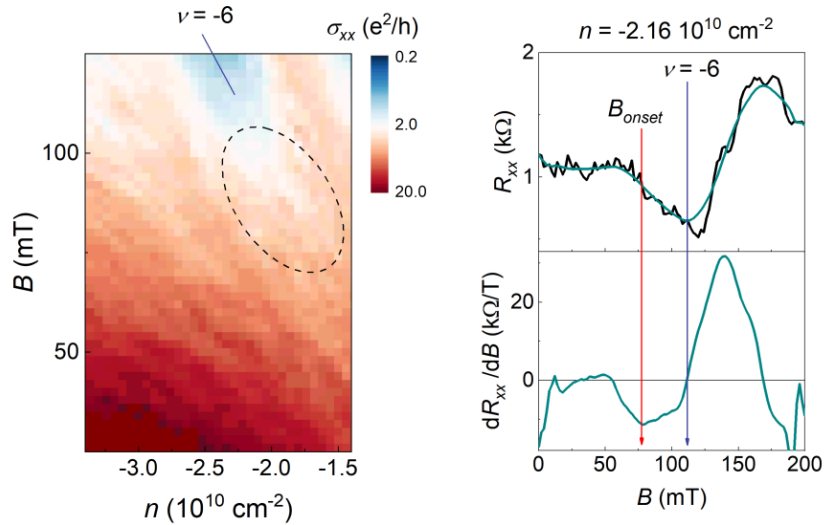

**Figure S6.** Left: zoom-in of main text Figure 4a, highlighting the onset of the $\nu$ = -6 oscillation. Right: procedure for the determination of the onset field. Top panel: longitudinal resistance as a function of the magnetic field at density -2.16 × 10$^{10}$ cm$^{-2}$. The black (dark cyan) line are experimental data (smoothed curve by adjacent averaging). Lower panel: first derivative of the longitudinal resistance (calculated from the smoothed curve). The $v = -6$ minimum corresponds to a zero in the derivative, while the onset field is identified at the largest negative value of $dR_{xx}/dB$ preceding the minimum.

**Imperfect quantization in the FQH regime**

As reported in the main text, we do not measure a precisely quantized value of $\sigma_{xy}$ in correspondence of $v = -1/3$. Since field-symmetrized data at ±12 T do not provide a substantial improvement, we exclude contacts' misalignment as the origin of this discrepancy. More likely, the lack of quantization is due to two factors related to the simple Hall bar geometry. The first one is the roughness of the etch-defined edges, which, despite common knowledge on topologically-protected phases, can strongly influence the edge states' transport: edge-free geometry [35, S5] or electrostatically-defined channels [S6] can mitigate this issue. Additionally, and possibly more importantly, perfect equilibration of the edge states at the metallic leads is required in order to observe quantization of $\sigma_{xy}$. Optimal equilibration is difficult to achieve in devices such ours, where a single global back-gate controls the carrier density both in the channel and in the contact regions. Maher *et al.* [S7] showed that a local bottom gate geometry allows tuning the sample to low filling factors, while keeping highly doped and efficient contacts via the Si back-gate. This strategy results in precise quantization in the fractional quantum Hall regime, absent otherwise. Implementing these advancements in the device fabrication should facilitate a complete establishment of FQH in CVD-G, of which our current data provide preliminary evidence.

**Additional References**